\documentclass[fleqn,usenatbib]{mnras}

\usepackage{newtxtext,newtxmath}
\usepackage[T1]{fontenc}

\usepackage{graphicx}	
\usepackage{bm}
\usepackage{hyperref}
\hypersetup{breaklinks=true}

\DeclareRobustCommand{\VAN}[3]{#2}
\let\VANthebibliography\thebibliography
\def\thebibliography{\DeclareRobustCommand{\VAN}[3]{##3}\VANthebibliography}

\title[Separatrix Divergence]{Separatrix Divergence of Stellar Streams in Galactic Potentials}

\author[T. D. Yavetz et al.]{Tomer D. Yavetz,$^{1}$\thanks{E-mail: t.yavetz@columbia.edu (TDY)}
Kathryn V. Johnston,$^{1,2}$
Sarah Pearson,$^{2}$
Adrian M. Price-Whelan,$^{2,3}$
\newauthor
Martin D. Weinberg$^{4}$
\\
$^{1}$Department of Astronomy, Columbia University, 550 West 120th Street, New York, NY 10027, USA\\
$^{2}$Center for Computational Astrophysics, Flatiron Institute, New York, NY 10010, USA\\
$^{3}$Department of Astrophysical Sciences, Princeton University, 4 Ivy Lane, Princeton, NJ 08544, USA\\
$^{4}$Department of Astronomy, University of Massachusetts at Amherst, 710 N. Pleasant St., Amherst, MA 01003
}

\date{Accepted XXX. Received YYY; in original form ZZZ}

\pubyear{2020}

\begin{document}
\label{firstpage}
\pagerange{\pageref{firstpage}--\pageref{lastpage}}
\maketitle

\begin{abstract}
Flattened axisymmetric galactic potentials are known to host minor orbit families surrounding orbits with commensurable frequencies. The behavior of orbits that belong to these orbit families is fundamentally different than that of typical orbits with non-commensurable frequencies. We investigate the evolution of stellar streams on orbits near the boundaries between orbit families (separatrices) in a flattened axisymmetric potential. We demonstrate that the separatrix divides these streams into two groups of stars that belong to two different orbit families, and that as a result, these streams diffuse more rapidly than streams that evolve elsewhere in the potential. We utilize Hamiltonian perturbation theory to estimate both the timescale of this effect and the likelihood of a stream evolving close enough to a separatrix to be affected by it. We analyze two prior reports of stream-fanning in simulations with triaxial potentials, and conclude that at least one of them is caused by separatrix divergence. These results lay the foundation for a method of mapping the orbit families of galactic potentials using the morphology of stellar streams. Comparing these predictions with the currently known distribution of streams in the Milky Way presents a new way of constraining the shape of our Galaxy's potential and distribution of dark matter.
\end{abstract}




\section{Introduction}
\label{sec:intro}

In the standard $\Lambda$CDM picture of the Universe, galaxies form at the centers of dark matter haloes. The emerging picture from simulations of disk galaxies like the Milky Way suggests that the outer regions of dark matter haloes exhibit triaxial configurations, while the baryonic components cause the inner parts of the haloes to take on more spherical or axisymmetric (prolate/oblate) configurations \citep{Dubinski1994,Debattista2008,Prada2019}. In the context of these simulations, an observational diagnostic that could assist in characterizing the shape of the Milky Way's halo at different radii would be of high value.

One distinguishing factor between haloes of different shapes is the orbit families that they support. A spherical NFW halo, for example, supports only one family of orbits -- the \textit{loop} orbit family. A triaxial St\"ackel potential, on the other hand, supports three separate \textit{tube} orbit families, as well as the \textit{box} orbit family. Other non-axisymmetric configurations can support any number of additional minor orbit families around closed orbits defined by a commensurability (also referred to as \textit{resonant} orbits). Finally, most potentials are also expected to host a subset of \textit{irregular} or \textit{chaotic} orbits, though the significance of this subset varies widely from potential to potential. Observational evidence indicating whether and to what extent each of these orbit families is populated in our Galaxy would have direct implications on our understanding of the shape of the dark matter halo \citep{Valluri2012,Rojas-Nino2012}. Such observational evidence has unfortunately remained elusive, and the application of orbit family classification has yet to yield meaningful constraints on the shape of our Galactic halo.

A more common approach for constraining the shape of our Galaxy's dark matter halo relies on the morphology of tidally disrupted Galactic satellites \citep{Johnston1999}. The tidal debris from these satellites is expected to populate the halo of our Galaxy with streams of stars that approximately trace the orbital path of the stream's progenitor \citep{Johnston1998,Helmi1999,Hendel2015,Erkal2016}. The orbits traced by streams in the Milky Way have been used to estimate both the Galaxy's mass and its shape (see, e.g., \citealt{Law2010,Vera-Ciro2013,Gibbons2014}). However, there are known complications with this orbit-fitting technique \citep{Sanders2013} and applying it to a single stream leads to a variety of uncertainties and biases \citep{Bonaca2014}. The resulting fits of the Milky Way's potential tend to vary widely, and degeneracies between the halo's shape and its density profile have led to difficulties in obtaining conclusive constraints (see, e.g., \citealt{Ibata2013}). One promising approach that may negate many of these issues involves combining information from multiple streams \citep{Bonaca2014,Reino2019}.

Recently, it has been shown that the morphology of stellar streams in certain regions of triaxial haloes can deviate greatly from the expected dynamically cold 1D morphology within just a few orbital times \citep{Pearson2015,Fardal2015,Price-Whelan2016,Mestre2019}. On the one hand, a fanned-out stream such as the ones described in these works cannot be used very effectively in the orbit fitting techniques cited above. On the other hand, the mere possibility of producing such effects in certain potentials can be used to effectively rule out certain halo geometries \citep{Pearson2015}. Understanding the mechanism (or mechanisms) for stream-fanning, the types of potentials that cause stream-fanning, and the locations and timescales of stream-fanning in those potentials may therefore unlock a new avenue for constraining the shape of the Galaxy's dark matter halo.

Under certain circumstances, stream-fanning can arise when the progenitor is on a strongly chaotic orbit, as shown in \citet{Fardal2015}, \citet{Mestre2019}, and for one of the cases discussed in \citet{Price-Whelan2016}. Yet if chaos is the main mechanism for inducing stream-fanning, one may question the feasibility of relying on it in order to constrain the shape of our Galaxy, given that only a very small subset of galactic orbits are expected to exhibit detectable effects of chaos on timescales shorter than the age of the Universe \citep{Goodman1981,Maffione2018}.

\citet{Pearson2015} and \citet{Price-Whelan2016} both found that stream-fanning can sometimes happen sooner than the onset of chaos (as predicted by the Lyapunov times of the progenitor orbits). This brings into question whether chaos is responsible for the fanning-out of the streams in these cases, while at the same time reviving the idea that stream-fanning may prove useful in constraining the shape of the Galaxy after all.

In what follows, we describe a different mechanism that can cause stellar streams to fan-out after only a few tens of orbital periods (i.e., well within the age of the Universe). We refer to this mechanism as \textit{separatrix divergence}, and argue that it can cause stream-fanning in any potential that hosts multiple orbit families, when the progenitor's orbit is close to the boundary between two orbit families. This includes a much broader subset of galactic potentials, including many axisymmetric potentials in addition to triaxial ones. The separatrices that define these boundaries represent a discontinuity in the orbital structure of the potential, and the fundamental properties of orbits on either side of the separatrix can differ significantly. When the progenitor orbit of a stellar stream is close enough to a separatrix between two orbit families, the ensemble of particles that the stream consists of `straddles' the separatrix, leading to the rapid divergence of the stream in phase-space.

The separatrices between orbit families are also the locations where stochasticity arises in some potentials, and these effects are indeed directly linked. However, we demonstrate in what follows that \textit{separatrix divergence} is a more general phenomenon that does not require the existence of a band of chaotic orbits to cause significant disruptions to the morphological evolution of a stellar stream.

This paper is organized as follows: we begin in \S\ref{sec:review} with a review of key theoretical concepts including fundamental properties of orbit families and Hamiltonian perturbation theory. \S\ref{sec:ensembles} describes our approach for modelling stellar streams as ensembles of unbound test particles. In \S\ref{sec:axisym} we demonstrate the effects of \textit{separatrix divergence} in an axisymmetric potential and compare the theoretically derived quantities from \S\ref{sec:review} to the results from test particle simulations. \S\ref{sec:triaxial} is devoted to determining whether the fanned-out stream morphologies investigated in \citet{Price-Whelan2016} and \citet{Pearson2015} are the result of \textit{separatrix divergence}. We discuss the limitations and future prospects of using \textit{separatrix divergence} to constrain galactic potentials in \S\ref{sec:discussion}, and present our main conclusions in \S\ref{sec:conclusion}.

Throughout sections \ref{sec:review} and \ref{sec:axisym} we use galactocentric polar coordinates $(R, z, \phi)$ to investigate \textit{separatrix divergence} in an axisymmetric Miyamoto-Nagai potential \citep{Miyamoto1975}:
\begin{equation}
\label{eq:potMN}
\Phi = -\frac{GM}{\sqrt{R^2 + (a + \sqrt{z^2 + b^2})^2}} \ .
\end{equation}
For simplicity we choose to work with dimensionless units ($GM = a = 1$), and we set $b/a = 0.45$.

In \S\ref{sec:triaxial} we turn our attention to \textit{separatrix divergence} in the two triaxial potentials considered in \citet{Price-Whelan2016} and \citet{Pearson2015}: a triaxial Lee-Suto \citep{Lee} potential with $b/a = 0.77$ and $c/a = 0.55$, and the multi-component \citet{Law2010} fit to the Milky Way's potential that includes a triaxial halo.

\section{Review of Terminology and Physical Intuition}
\label{sec:review}

In this section we outline several foundational concepts in orbital dynamics upon which we rely throughout the rest of this work. We begin in \S\ref{subsec:orbits} with a brief discussion of galactic orbits and the fundamental quantities used to describe them. In \S\ref{subsec:orbit_families} we review the orbit families that arise in galactic potentials and the characteristics that can be used to differentiate between them. Lastly, we devote \S\ref{subsec:perturbation_theory} to deriving several quantities of interest from Hamiltonian perturbation theory that will be used to estimate the relevant time- and mass-scales for \textit{separatrix divergence}.

\subsection{Fundamental Properties of Orbits and Canonical Coordinates}
\label{subsec:orbits}

Systems of regular orbits are often described in terms of action-angle variables ($\bm{J}, \bm{\theta}$), the natural set of canonical coordinates that result from separating the Hamilton-Jacobi equation. By construction, Hamilton's equations then reduce to linear increasing angle values ($\bm{\theta} = \bm{\theta_0} + \bm{\Omega}t$, where $\bm{\Omega} = \partial H / \partial \bm{J}$). While the general orbit fills the orbital torus, a measure zero set with rational frequency ratios describe closed orbits (i.e., $\bm{n} \cdot \bm{\Omega} = 0$ for a set of finite integers $\bm{n} \neq 0$). It is customary to refer to an orbit with fundamental frequencies related by $\Omega_1:\Omega_2 = m:n$ for finite integers $m$ and $n$ as the $m$:$n$ commensurable orbit. 

Some $N$-dimensional potentials also support \textit{irregular} or \textit{chaotic} orbits that do not conserve $N$ integrals of motion. These orbits cannot be described accurately by action-angle coordinates, though in some mildly chaotic cases it is possible to approximate long segments of the orbit with fixed actions.

There are a variety of numerical methods for converting the phase-space coordinates of an orbit to action-angle coordinates. In what follows, we mainly rely on the \citet{Sanders2014} method, using the publicly available \url{Gala} package \citep{Price-whelan2017}. In some instances, we use spectral analysis to evaluate the fundamental frequencies instead of performing the full transformation to action-angle coordinates. This is particularly convenient for rapidly detecting orbits with commensurable fundamental frequencies. For this purpose we utilize \url{SuperFreq} \citep{Price-Whelan2015}, which follows the Numerical Approximation of Fundamental Frequencies (NAFF) method introduced in \citet{Laskar1993}. For more information, see appendix B in \citet{Price-Whelan2016}.

Finally, it is also necessary to perform the reverse conversion from action-angle coordinates to phase space coordinates in order to understand the resonant trapping of orbits through Hamiltonian perturbation theory. For this we follow the approach outlined in \citet{Mcgill1990}, \citet{Binney1993}, and \citet{Kaasalainen1994b}, which is also conveniently available in the \url{TorusMapper} package \citep{Binney2016a} through \url{Galpy} \citep{Bovy2014a} or \url{AGAMA} \citep{Vasiliev2019}. \citet{Binney2016b} discusses specifically how the code has been extended to model resonant trapping, and demonstrates its use for the same potential we focus on throughout much of what follows.

\subsection{Orbit Families}
\label{subsec:orbit_families}

Regular galactic orbits can be classified into a set of orbit families. Each orbit family shares certain characteristics, and is \textit{parented} by a closed orbit. A convenient approach to classifying orbits into orbit families relies on the spectral analysis of orbits \citep{Binney1982}. In what follows, we adopt much of the language of the spectral classification methodology laid out in \citet{Carpintero1998}.

Most commonly used 2-D potentials have two orbit families: the \textit{box} orbit family and the \textit{loop} orbit family. The box orbit family is parented by an axial orbit that moves along the long axis of the potential. The loop orbit family is parented by a closed loop orbit, and orbits belonging to this orbit family maintain their sense of revolution around the center of the potential and conserve angular momentum.

Some 2-D potentials also support minor orbit families that are embedded within the major orbit families. The parent orbits of these minor orbit families are closed orbits that are defined by commensurable fundamental frequencies. All non-parent members of the various minor orbit families will have an additional fundamental frequency that defines the libration of the orbit around the parent orbit. The orbits belonging to these minor orbit families are sometimes referred to as orbits that are `trapped' by a resonance.

In 3-D potentials the situation is mostly analogous to the 2-D cases described above. For non-axisymmetric (triaxial) galactic potentials, the loop orbit family from the 2-D case splits into three \textit{tube} orbit families: the short-axis tube orbit family and two separate long-axis tube orbit families, known as the inner- and the outer long axis tube orbit families \citep{Binney2008}. In addition to these three tube orbit families and the box orbit family, boxlets and looplets defined by commensurable frequencies can also parent minor orbit families in 3-D (for a more complete treatment of orbit family classification in 3-D, see \citealt{Carpintero1998}).

\begin{figure}
\includegraphics[width=\columnwidth]{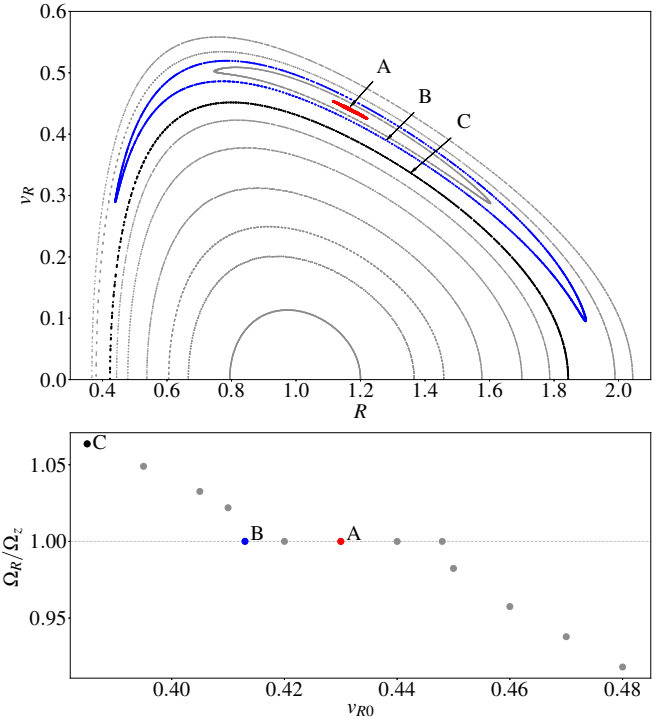}
\caption{\textit{Upper panel}: A surface of section for the Miyamoto-Nagai potential described in equation \ref{eq:potMN}. The contours are obtained by initializing a series of orbits with $E = -0.39$ and $L_z = \sqrt{0.075}$, integrating them, and logging the values of $R$ and $v_R$ every time the orbit crosses the $z = 0$ plane with positive $v_z$. The majority of contours describe typical tube orbits, while the minor orbit family defined by the $\Omega_R:\Omega_z = 1:1$ commensurability can be seen near the top of the figure. Three orbits of interest are marked in the figure: orbit A is very close to the parent orbit of the minor orbit family, orbit B is still a member of the minor orbit family but is adjacent to the separatrix, and orbit C is a tube orbit on the other side of the separatrix. \textit{Lower panel}: The ratio $\Omega_R / \Omega_z$ for a series of orbits around the 1:1 commensurability (specified by the initial value of $v_R$ for each orbit, while $R_0 = 1.2$ and $z_0 = 0$ for all of the orbits). The discontinuity in frequency space near the boundary between the orbit families can be seen clearly in this plot. The five points in the middle represent members of the minor orbit family for which the ratio of the frequencies is 1. The frequency ratios of the three labeled orbits in the upper panel are also indicated in the lower panel.}
\label{fig:01}
\end{figure}

In what follows, we focus our attention to the 3-D flattened axisymmetric potential specified in equation \ref{eq:potMN}. This is a convenient choice for a couple of reasons: first, due to its axisymmetric nature, it is possible to easily visualize orbits in this potential using a surface of section, as shown in the upper panel of Figure \ref{fig:01}. Second, this potential hosts a prominent minor orbit family defined by commensurable radial and vertical frequencies (i.e., $\Omega_R:\Omega_z = 1:1$). This minor orbit family, informally known as the `banana' or `saucer' orbit family due to the characteristic shape of its orbits, appears as an island of contours near the top of the surface of section. The border between the major and the minor orbit families represents a sharp frequency discontinuity (or, equivalently, a discontinuity in the mapping from physical to action space). Aside from the appearance of an island in the surface of section, this discontinuity is also readily identifiable through the spectral analysis of adjacent orbits at the transition between the orbit families, as shown in the lower panel of Figure \ref{fig:01}.

The behavior of orbits in and near minor orbit families such as this one has been studied extensively through the application of perturbation theory. We devote the next section to a review of the derivations that enable us to accurately describe the orbits belonging to this minor orbit family.

\subsection{Secular Perturbation Theory}
\label{subsec:perturbation_theory}

As discussed in \citet{Lichtenberg1992} and shown in detail for the Miyamoto-Nagai potential in \citet{Binney1993}, the behavior of orbits in the vicinity of an orbit with commensurable frequencies can be analyzed through Hamiltonian perturbation theory. This necessitates a canonical coordinate transformation into a new set of actions and angles $(\bm{J}', \bm{\theta}')$ that describe the libration of an orbit around the parent orbit of that family. This motion can be described using a Hamiltonian of the form:
\begin{equation}
\label{eq:hamiltonian}
H(\bm{J}', \bm{\theta}') = H_0(\bm{J}') + \delta H(\bm{J}', \bm{\theta}') \ ,
\end{equation}
where $H$ is the original Hamiltonian (now defined by the new actions and angles), $H_0$ is an integrable Hamiltonian that is only a function of the new actions $\bm{J}'$, and $\delta H$ is a small perturbation compared to $H_0$ that is periodic with respect to the new angles $\bm{\theta}'$.

For the parent orbit of the minor orbit family, $\bm{n} \cdot \bm{\Omega} = 0$, so in the vicinity of this orbit, the linear combination of the angle variables $\theta_s = \bm{n} \cdot \bm{\theta}$ will evolve slowly. We are interested in transforming to the new set of actions and angles that include the \textit{slow angle} $\theta_s$ and its conjugate action coordinate $J_s$ (the other action and angle variables in this new coordinate system will be referred to as the \textit{fast} actions and angles, with the subscript $f$). This requires a generating function $W(\bm{J}', \bm{\theta}, t)$ that satisfies:
\begin{equation}
\label{eq:genfunc1}
\bm{\theta}' = \frac{\partial W}{\partial \bm{J}'} \quad \text{and} \quad \bm{J} = \frac{\partial W}{\partial \bm{\theta}} \ .
\end{equation}

The first of these suggests the following form of $W$:
\begin{equation}
\label{eq:genfunc2}
W = \theta_1J_{f1} + \theta_2J_{f2} + (\bm{n}\cdot\bm{\theta})J_s ,
\end{equation}
which yields $\theta_{f1} = \theta_1$, $\theta_{f2} = \theta_2$, and $\theta_s = \bm{n} \cdot \bm{\theta}$. The second equation in \ref{eq:genfunc1} specifies:
\begin{equation}
\label{eq:slow_actions}
J_1 = \frac{\partial W}{\partial \theta_1} = J_{f1} + N_1J_s \ ; \quad J_2 = J_{f2} + N_2J_s \ ; \quad J_3 = N_3J_s \ .
\end{equation}

We can use the new angle variables to take the Fourier expansion of the Hamiltonian perturbation in equation \ref{eq:hamiltonian}:
\begin{equation}
\label{eq:fourier_hamiltonian1}
\delta H = \sum_{\bm{n}\neq0} H_{\bm{n}}(\bm{J}')\exp(i\bm{n}\cdot\bm{\theta}') \ .
\end{equation}

The rapid evolution of the fast angles compared to $\theta_s$ in the vicinity of the commensurable orbit allows us to treat $J_{f1}$ and $J_{f2}$ as adiabatic invariants and average the equation above over $\theta_{f1}$ and $\theta_{f2}$ to obtain:
\begin{equation}
\nonumber
\hat{H}(\theta_s, \bm{J}') = H_0(J_s) + \sum_{m\neq0}H_m(J_s)\exp(im\theta_s)
\end{equation}
\begin{equation}
\label{eq:fourier_hamiltonian2}
 = H_0(J_s) + 2 \sum_{m\neq0}H_m(J_s)\cos(m\theta_s) \ ,
\end{equation}
in which we have also phase-shifted $\theta_s$ to simplify the equation. Due to the symmetry of our chosen potential with respect to $z=0$, the Fourier coefficients $H_m$ vanish for all odd $m$. Furthermore, the higher order coefficients decrease rapidly, leaving $m=2$ as the dominant term \citep{Binney1993}. Taking these approximations into account and expanding $H_0$ to second order in $\delta J_s = J_s - J_{s0}$ (where $J_{s0}$ is the slow action of the parent orbit) yields the pendulum equation \citep{Chirikov}\footnote{\citet{Kaasalainen1994} demonstrated how a modified pendulum equation that retains higher order terms can describe the motion near the commensurable orbit to a higher level of accuracy, but in this work we find that the approximate treatment of resonances shown above is sufficient to capture the key elements of the behavior of streams in the vicinity of separatrices.}:
\begin{equation}
\label{eq:pendulum}
\Delta \hat{H} = \frac{1}{2}G\times(\delta J_s)^2 - F\cos(m\theta_s) \ ,
\end{equation}
with:
\begin{equation}
\label{eq:GandF}
G \equiv \frac{\partial^2 H_0}{\partial J_s^2}\bigg|_{J_{s0}} \quad \text{and} \quad F \equiv -2H_2(J_{s0}) \ .
\end{equation}

Orbits with $\Delta \Hat{H} < 0$ are the \textit{librating} orbits that make up the minor orbit family, whereas orbits with $\Delta \Hat{H} > 0$ are \textit{circulating} orbits that do not belong to the minor orbit family. The extent of the minor orbit family in action space can thus be estimated through the half-width $\delta J_s$ for the \textit{separatrix orbit} (i.e., when $\Delta \hat{H}$ in equation \ref{eq:pendulum} is set to zero):
\begin{equation}
\label{eq:libration_action}
\delta J_{s,\text{max}} \simeq \sqrt{\frac{2F}{G}} \ .
\end{equation}

The libration frequency near the parent orbit can be obtained by applying Hamilton's equations to equation \ref{eq:fourier_hamiltonian2} and differentiating $\dot{\theta}_s$ with respect to time:
\begin{equation}
\label{eq:libration_freq}
\omega_l \simeq \sqrt{GF} \ .
\end{equation}
Just like a pendulum, the libration frequency is not the same for every orbit in the minor orbit family and it shrinks to zero at the separatrix. Nonetheless, the libration period ($T_l = 2\pi / \omega_l$) still provides a useful timescale for understanding the behavior of orbits that belong to the minor orbit family.

Applying these equations to the $\Omega_R : \Omega_z = 1:1$ commensurability in the potential in equation \ref{eq:potMN} yields $\delta J_{s\text{,max}} = 0.015$ and $T_l \sim 36$ orbital periods.

\begin{figure*}
\includegraphics[width=0.9\textwidth]{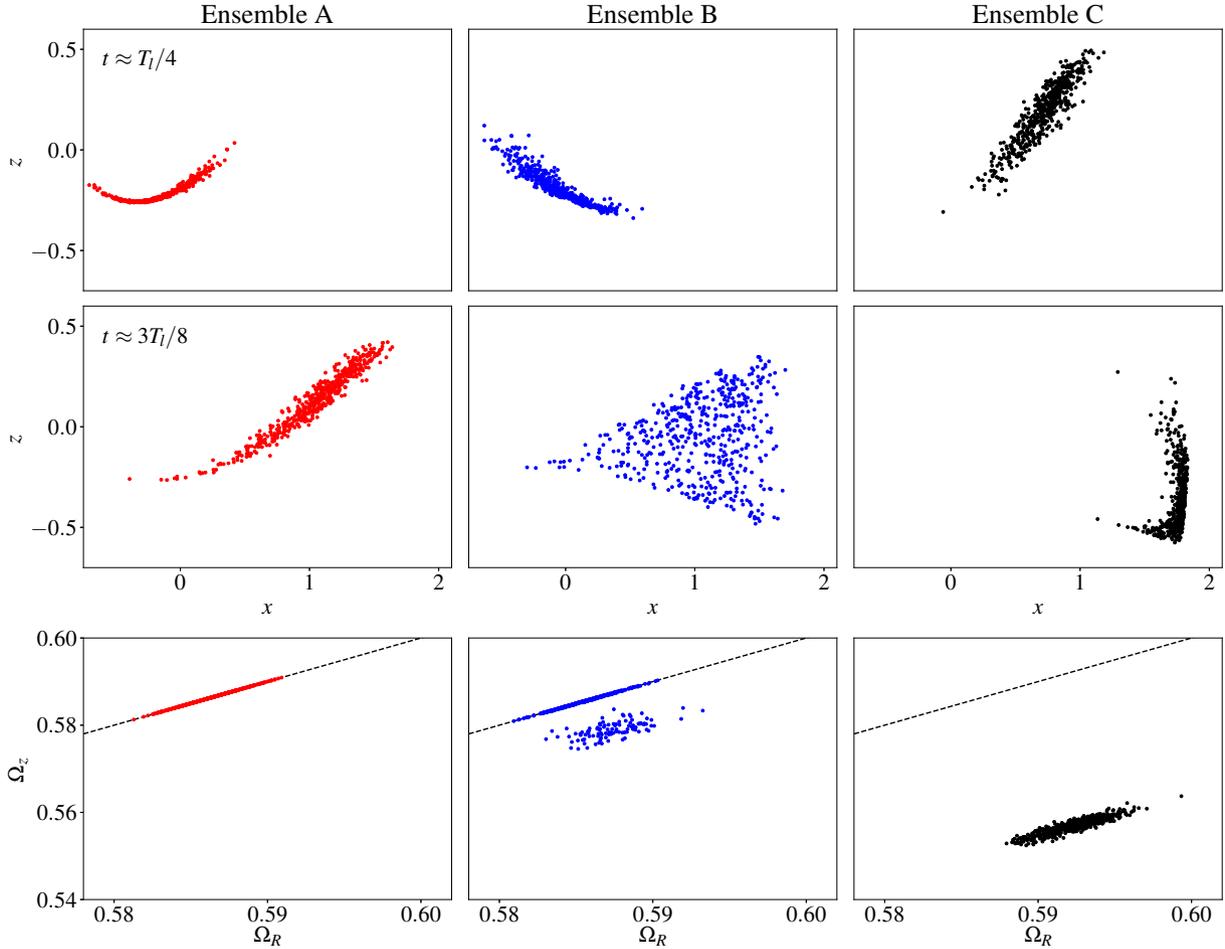}
\caption{\textit{Upper rows}: the morphology of three streams with $m/M = 10^{-7}$ (modelled as unbound ensembles following equation \ref{eq:action-spread} around the three orbits indicated in Figure \ref{fig:01}) in two different snapshots. After a period in which the three streams seem to evolve similarly, ensemble B abruptly becomes more diffuse and fanned-out. \textit{Lower row}: the radial and vertical fundamental frequencies of the particles in the upper panels. The 1:1 commensurability is marked by a dotted line. Ensembles A and C form cohesive groups in frequency space, while Ensemble B is split into two distinct groupings, one of which is trapped at the commensurability.}
\label{fig:02}
\end{figure*}

\section{Modelling Stellar Streams as Unbound Stellar Ensembles}
\label{sec:ensembles}

The ultimate aim of this work is to apply the insights about individual orbits near separatrices from the previous section to the morphology and evolution of stellar streams. We therefore turn our attention in this chapter to a few general properties of stellar streams that will allow us to relate the quantities and timescales from the previous section to the expected morphology of a stream evolving near a separatrix.. 

The observed debris structures with stream-like morphologies that orbit our Galaxy are thought to form from the tidal disruption of bound stellar ensembles (such as globular clusters of dwarf galaxies). While the conditions for a particle to become unbound from its progenitor are non-trivial, once the debris has been unbound and stripped from its progenitor, it is well-modelled by a collection of non-gravitating test particles with a small spread in initial conditions, energy, and angular momentum \citep{Johnston1998,Helmi1999,Kupper2012,Fardal2015}. The spread in these quantities is centered around the energy and angular momentum of the \textit{progenitor orbit} -- the orbit of the progenitor object from which the unbound ensemble originated. 

The spread of an unbound ensemble's initial conditions in 6-D phase-space is related to the tidal scale $(m/M)^{1/3}$ (e.g., \citealt{Johnston1998}), where $m$ is the total mass of the disrupted object and $M$ is the mass of the disrupting object enclosed within the orbit of the disrupted object. To model stellar ensembles in a galactic potential without resorting to computationally expensive N-body codes, we assume that debris structures are well-approximated by ensembles of massless test particles whose initial conditions follow a normal distribution centered around the initial conditions of the progenitor orbit (as specified in equations (13)-(16) in \citealt{Price-Whelan2016}, among others).

The scale of this spread can also be related directly to the spread of the particles' orbits in action space:
\begin{equation}
\label{eq:action-spread}
\frac{\Delta J_i}{J_i}\sim \bigg(\frac{m}{M}\bigg)^{1/3}\ ,
\end{equation}
and since the actions and the fundamental frequencies are directly related to each other, we can also relate this scale to the spread in frequencies through $\Delta J$:
\begin{equation}
\label{eq:hessian}
\Delta\Omega_i = \sum_{j=1}^3\Delta J_j\frac{\partial^2H}{\partial J_i\partial J_j}\ ,
\end{equation}
where the last term is the Hessian of the Hamiltonian \citep{Tremaine1999,Sanders2013}. In most cases, the local Hessian is dominated by a single eigenvalue, and consequentially equation \ref{eq:hessian} requires that the distribution of fundamental frequencies be close to one-dimensional.

The 1-D spread in frequency space translates into the 1-D evolution of the particles. The dominant eigenvalue leads to a characteristic frequency width that describes the time required for leading particles to overtake trailing particles:
\begin{equation}
\label{eq:phase_mixing}
T_\text{pm} \sim \frac{2\pi}{\Delta\Omega} \ .
\end{equation}

For a typical globular cluster ($m\approx10^4-10^5\ M_\odot$) orbiting the Milky Way, the spread in fundamental frequencies around the progenitor orbit is typically around 0.1\%-1\%, and since this scale is inversely related to the phase-mixing timescale, it can take a stream like this hundreds of orbital periods to fully phase mix. As the stream phase mixes, its density decreases gradually. \citet{Helmi1999} found that the density of streams in axisymmetric potentials decreases as $t^{-3}$ in axisymmetric potentials (or as $t^{-2}$ in spherical potentials).

\begin{figure*}
\includegraphics[width=0.9\textwidth]{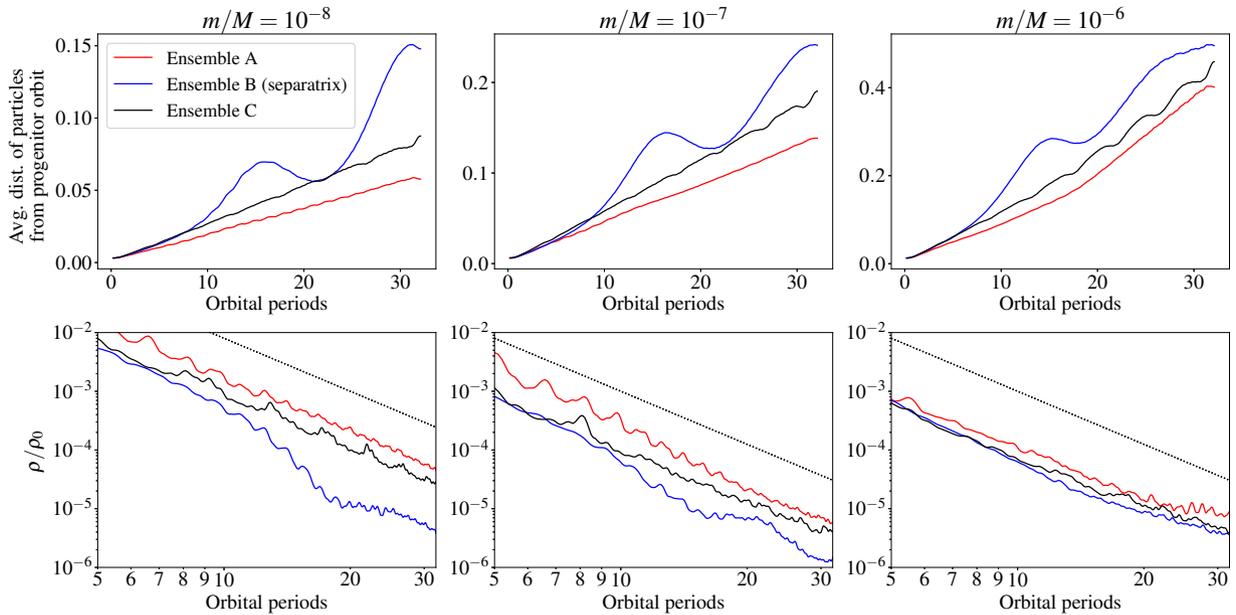} 
\caption{The dependence of \textit{separatrix divergence} on the mass of the progenitor. Each column depicts the morphological evolution of three ensembles whose progenitors are initialized on orbits A, B, and C from Figure \ref{fig:01}. The mass scale of the progenitors in each column is specified at the top. \textit{Upper row}: the mean distance of the ensemble's particles from the progenitor's orbit (smoothed over a few orbital periods to remove short-term oscillations between peri- and apocenter). Unlike ensembles A and C, which exhibit linear growth in these plots, ensemble B displays an unmistakable long-term oscillatory behavior in addition to the linear growth of the other ensembles. The period of this oscillation matches the predicted libration period from secular perturbation theory. \textit{Lower row}: the mean density of each ensemble over time divided by the ensemble's initial density. The ensemble density is calculated by implementing a kernel density estimation routine with an adaptive Epanechnikov kernel, and smoothed over a few orbital periods to remove short-term density fluctuations between peri- and apocenter. The densities of ensembles A and C roughly follow the expected $t^{-3}$ power law (plotted for reference as a black dotted line). The density of ensemble B follows the same power law for most of its evolution, but undergoes periods of accelerated diffusion corresponding to when the progenitor orbit approaches the hyperbolic fixed points (at the same times at which the blue lines in the upper panels diverge from the other ensembles). The significance of the effect diminishes as the progenitor size grows: the $m/M=10^{-8}$ ensemble becomes more diffuse than the nearby ensembles by over an order of magnitude within 15 orbital periods, whereas for the $m/M=10^{-7}$ ensemble the difference is closer to a factor of $\sim5$, and for the $m/M=10^{-6}$ ensemble the effect is no longer strong enough to cause any noticeable differences compared to nearby streams that do not experience \textit{separatrix divergence}. This, too, is in agreement with the predictions from the comparison of equations \ref{eq:libration_action} and \ref{eq:action-spread}.}
\label{fig:03}
\end{figure*}

\section{Results: Exploration of \textit{Separatrix Divergence} in Axisymmetric Potentials}
\label{sec:axisym}

We now turn to applying the machinery from \S\ref{sec:review} and \S\ref{sec:ensembles} to investigate the evolution of streams whose progenitor orbit is close to a separatrix.

If the progenitor is close enough to the separatrix relative to the stream's intrinsic spread of initial conditions, some particles will fall on the other side of the separatrix and evolve on orbits belonging to a different orbit family. These particles' orbits should therefore diverge from the progenitor orbit -- and the orbits of the rest of the stream's particles -- after no more than $T_l / 2 \simeq 18$ orbital periods. If this occurs before the typical time it takes for the ensemble to become phase-mixed, we argue that the stream evolving close to the separatrix should become considerably more diffuse and less stream-like than other ensembles evolving in the same potential on orbits that do not bring them near a separatrix.

To test this hypothesis, we model test-particle streams following the prescription in \S\ref{sec:ensembles} whose progenitors are placed on the orbits marked A, B, and C in Figure \ref{fig:01}. For each orbit, we initiate three unbound test particle ensembles whose initial distribution is governed by mass ratios ranging between $m/M = 10^{-8}$ and $m/M = 10^{-6}$ (for galactic masses this is roughly equivalent to globular clusters with masses between $10^4$ M$_\odot$ and $10^6$ M$_\odot$). In each case, ensemble A is embedded deep within the minor orbit family shown in Figure \ref{fig:01} such that all of the ensemble's particles are on commensurable orbits. Ensemble C is likewise so far outside the minor orbit family that none of its particles are on orbits that belong to the minor orbit family. Ensemble B, however, is initiated very close to the separatrix, and its particles end up divided between the two orbit families.

The morphology of ensembles A, B, and C for the middle mass ratio ($m/M = 10^{-7}$) is shown at two different times in the top two rows of Figure \ref{fig:02}. Over the first $T_l / 4$ ($\approx9$ orbital periods) all three streams maintain a relatively thin and stream-like morphology, whereas just 5 orbital periods later stream B has become significantly more diffuse and fanned-out compared to the other two.

The bottom row of Figure \ref{fig:02} captures the same three streams in frequency space. The fundamental frequencies of all of the particles that belong to ensemble A are tightly clustered around the $\Omega_R : \Omega_z = 1:1$ line, indicating clearly that they all belong to the minor orbit family defined by the commensurability. Likewise, the particles that make up ensemble C all belong to the major tube orbit family and appear as a cohesive (albeit slightly less thin) cluster in frequency space. Contrary to the other two ensembles, ensemble B `straddles' the separatrix between the two orbit families, and as a result, some of its particles belong to the minor orbit family and exhibit the commensurability, while others belong to the major orbit family.

\begin{figure*}
\includegraphics[width=0.9\textwidth]{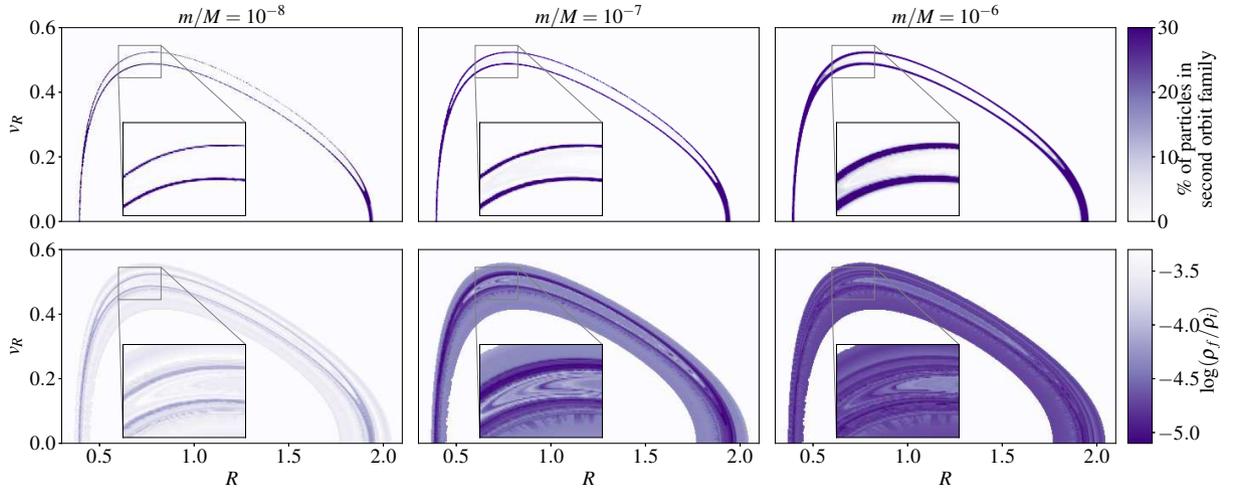} 
\caption{\textit{Upper row}: the percentage of ensemble particles that belong to a second orbit family for the same three mass scales analyzed in Figure \ref{fig:03}. Ensembles of 256 particles each are generated on a dense grid of progenitor orbits at each mass scale, and the fundamental frequencies of each particle are analyzed over $\sim100$ orbital periods (the exact same procedure used to produce the lower panels of Figure \ref{fig:02}). Each particle with $|\Omega_R / \Omega_z - 1| < 0.005$ was classified as belonging to the commensurable orbit family. The colorbar is truncated at $30\%$ to more clearly emphasize the width of the band that may experience \textit{separatrix divergence}. As the mass of the progenitor is increased, the width of the band around the separatrix grows. \textit{Lower row}: the simulated density of ensembles after $\sim15$ orbital periods on the same grid of initial conditions and the same three mass scales as the upper row (the grid extends between the orbits with $(x_0, v_{x0}) = (1.2, 0.35)$ and $(1.2, 0.48)$, and no comparable features appear in density space outside of this range). The affected band around the separatrix grows as in the upper panels, but as the progenitor mass increases, the difference in the density of streams affected by \textit{separatrix divergence} and those not affected shrinks. The overall significance of the effect appears to peak around $m/M = 10^{-7}$. Note also that the affected region around the separatrix in the lower panels is slightly wider than that predicted by the upper panels. This is another result of the pendulum-like behavior of orbits near the separatrix -- even if all of the particles belong to the same orbit family, some may be close enough to the separatrix such that their libration (or circulation) phase will evolve very slowly near the hyperbolic fixed point, causing the ensemble to spread out in angle-space much more rapidly than normal.}
\label{fig:04}
\end{figure*}

As mentioned above, our expectation is that the group of particles on librating orbits will diverge from the rest of the particles after no more than $T_l / 2$. Given the initial phase of the progenitor, the divergence actually occurs after only $T_l/4 \simeq 9$ orbital periods. This is in good agreement with the two snapshots shown in Figure \ref{fig:02}. For comparison, equation \ref{eq:phase_mixing} yields the rough timescale of $\sim$ 60 orbital periods for $m/M=10^{-7}$, after which the ensemble particles will overlap for the first time in phase space, ruling out that regular phase-mixing is to blame for the fanned-out morphology of ensemble B. We therefore conclude that the ensemble's proximity to the separatrix is the primary driver of its fanned-out morphology, and refer to the resulting behavior as \textit{separatrix divergence}.

Our findings are further reinforced in Figure \ref{fig:03}, in which the morphological evolution of the ensembles over time is displayed using both the mean distance of particles from the progenitor orbit (upper panels) and the mean density of the ensembles (lower panels). The middle column corresponds to the evolution of the three ensembles depicted in Figure \ref{fig:02}. Due to the expected stream-orbit misalignment \citep{Sanders2013}, the quantity in the upper panels should typically grow linearly with time, as it appears to do for both ensembles A and C, but not ensemble B. The oscillatory nature of the separatix ensembles in all three of these panels is a direct consequence of the pendulum-like motion near the separatrix: after the initial divergence, the librating and circulating orbits come back together when they next pass by the hyperbolic fixed point.

This oscillatory behavior can also be seen in the density evolution shown in the lower panels, as the separatrix ensemble alternates between periods of $t^{-3}$ diffusion (like the other two ensembles) and short periods of exponential diffusion. The exponential diffusion is in some ways reminiscent of the diffusion of ensembles that evolve along strongly chaotic orbits, except that for \textit{separatrix divergence} this only appears to occur over short, periodic bursts.

In that context, it is worth noting that in potentials that host stochastic (chaotic) orbits, the stochastic regions typically appear as layers that surround separatrices. However, our potential was specifically chosen to be one with little stochasticity, in order to easily differentiate between \textit{separatrix divergence} and stream-fanning due to chaos. Just to be sure, we calculate the Lyapunov times for a representative subset of 64 particles in ensemble B, and find that they are all greater than 200 orbital periods ($t_\text{lyap} = 1 / \lambda_\text{max}$ where $\lambda_\text{max}$ is the maximum Lyapunov exponent of the orbit).

As noted in \S\ref{sec:ensembles}, the initial spread in the actions of the ensemble of particles that represents the disrupted cluster is governed by the ratio of the progenitor's mass and that of the host galaxy ($\Delta J$ in equation \ref{eq:action-spread}). This spread can be directly compared to the width of the minor orbit family in action space ($\delta J_{s,\text{max}}$ in equation \ref{eq:libration_action}). For ensembles with $\Delta J < \delta J_{s,\text{max}}$, \textit{separatrix divergence} should cause a noticeable drop in density after the expected timescale, whereas when $\Delta J \gtrsim \delta J_{s,\text{max}}$, the initial spread in action space will be large enough to compete with the morphological effects of \textit{separatrix divergence}.

For the commensurability discussed here, $\delta J_{s,\text{max}} \simeq 0.015$. For the separatrix ensembles (initiated on orbit B) when $m/M=10^{-8}$ and $m/M=10^{-7}$, $\Delta J_R \sim 0.002$ and $\sim 0.005$, respectively, placing them safely in the regime in which \textit{separatrix divergence} should cause noticeable morphological effects to the stream. This appears to be the case based on Figure \ref{fig:03}. However, for the $m / M = 10^{-6}$ ensemble, $\Delta J_R \sim 0.01 \sim \delta J_{s,\text{max}}$, and indeed the morphological effects of \textit{separatrix divergence} for this mass scale are less pronounced in Figure \ref{fig:03}. Further increasing the progenitor mass would render the effects of \textit{separatrix divergence} near this minor orbit family largely inconsequential.

Having shown the effect on a few representative orbits, we now turn to a preliminary exploration of mapping the potential. Figure \ref{fig:04} is based on the surface of section shown in Figure \ref{fig:01}. Each pixel in these plots represents the initial conditions of the progenitor orbit of an unbound ensemble of test particles. The shade of each pixel in the upper panel plots is determined by the percentage of particles on orbits belonging to a second orbit family. For comparison, in the bottom panel each pixel is shaded by the mean density of the ensemble after $\sim15$ orbital periods (calculated using the same kernel density estimation method used for the lower panels of Figure \ref{fig:03}). The three columns correspond to the same three mass scales shown in the columns of Figure \ref{fig:03}.

As the mass of the progenitor grows with respect to the mass of the host potential, the footprint of the ensemble in initial condition space grows, too, increasing the region in which the ensemble's particles will fall into two unique orbit families (i.e., the shaded band around the separatrix in the upper panels grows as the mass ratio is increased from $m/M = 10^{-8}$ to $10^{-6}$). At the same time, the morphological effects of \textit{separatrix divergence} become less noticeable once the initial spread in actions competes with $\delta J_{s,\text{max}}$, as seen in the reduced contrast between the ensembles affected by the separatrix and those not affected in the bottom right panel of Figure \ref{fig:04}.

The similarities between the upper and lower panels of Figure \ref{fig:04} serve to highlight two key points:
\begin{enumerate}
	\item \textit{Separatrix divergence} is the most significant cause of stream-fanning over the timescale shown for a smooth potential such as this one, and its effects can be clearly distinguished from the gradual diffusion of a typical stream as it becomes phase-mixed, provided that the mass ratio of the progenitor and the host galaxy is below a certain threshold.
	\item Figure \ref{fig:04} can be used to map a potential in terms of initial conditions of progenitors that may experience \textit{separatrix divergence}. In particular, the importance of the effect will peak for certain progenitors -- those whose mass is large enough to cause their members to extend beyond a separatrix for a large subset of initial conditions, and for which \textit{separatrix divergence} causes a larger spread in actions than the original tidal disruption of the cluster.
\end{enumerate}

Only a small subset of initial conditions in the potential analyzed above would lead to a globular cluster experiencing \textit{separatrix divergence}. However, the example discussed in this chapter is merely meant to prove the existence of this effect and to demonstrate how it can be analyzed and studied. More complex potentials with additional components and fewer degrees of symmetry may host many additional minor orbit families, each forming another region in which stellar streams may experience \textit{separatrix divergence}. A full determination of the significance and detectability of \textit{separatrix divergence} within realistic models of the Milky Way is beyond the scope of this paper, but will be addressed in future work.

\section{Discussion of Prior Work}
\label{sec:triaxial}

In the previous section, we developed an intuition and a methodology for analyzing the effects of \textit{separatrix divergence}, and we now turn to examining whether \textit{separatrix divergence} may be the effect causing stream-fanning in previously studied simulations. We focus on two cases in this section: in \S\ref{subsec:APW15} we review one of the ensembles investigated in \citet{Price-Whelan2016}, which was simulated in a \citet{Lee} triaxial potential. Then, in \S\ref{subsec:SP15}, we examine a disrupted ensemble meant to reproduce the stellar stream Palomar 5 (Pal 5) in the \citet{Law2010} potential, which was shown to fan out in simulations produced by \citet{Pearson2015}.

Before describing these two cases in detail, we note several complications that make studying \textit{separatrix divergence} in triaxial potentials considerably more challenging:

\begin{enumerate}
	\item The orbital structures of triaxial potentials are harder to visualize, as it is no longer possible to produce surface of section plots, making it more difficult to identify orbit families and the boundaries between them (but see \citealt{Schwarzschild1993} and \citealt{Carpintero1998} for examples of how this can still be accomplished).
	\item From a computational standpoint, applying the perturbation theory calculations from \S\ref{subsec:perturbation_theory} to obtain analytical estimates of the libration time and the width of minor orbit families requires the more sophisticated approach described in \citet{Kaasalainen1994b} and \citet{Kaasalainen1994} which is not available as of the writing of this paper in any of the astrophysical codes mentioned in \S\ref{sec:review}.
	\item \textit{Resonance overlap} can occur frequently in many triaxial potentials, possibly causing a larger subset of orbits to behave stochastically. Since stochasticity and \textit{separatrix divergence} both occur near separatrices and both can lead to stream-fanning, we must take additional care to distinguish between the two effects.
\end{enumerate}

\subsection{An Ensemble Near a Commensurability in the Lee-Suto (2003) Triaxial Potential}
\label{subsec:APW15}

We begin with \textit{separatrix divergence} in the \citet{Lee} triaxial potential. \citet{Price-Whelan2016} compared four ensembles initialized and simulated using both the test particle assumptions described in \S\ref{sec:ensembles} and N-body simulations, obtaining similar results in both cases. Two of the ensembles diffused considerably faster than the other two: one was on a ``strongly chaotic'' orbit with a Lyapunov time of $\sim8$ orbital periods. The other fanned-out ensemble was initialized on an orbit described as ``weakly chaotic'' with a Lyapunov time >700 orbital periods. This final ensemble (labeled by the letter C in \citealt{Price-Whelan2016}) is the focus of our analysis in this section.

\begin{figure}
\includegraphics[width=\columnwidth]{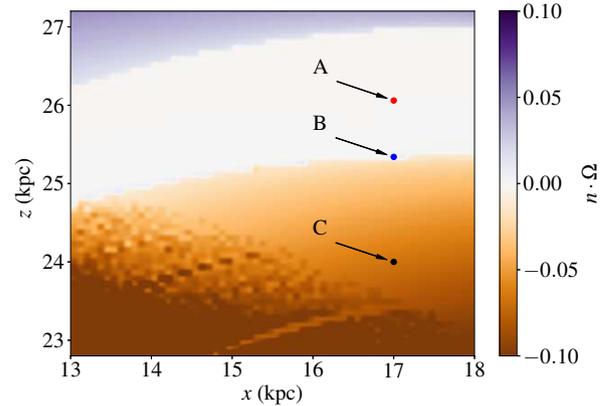} 
\caption{The value of $\bm{n}\cdot\bm{\Omega}$ for $\bm{n} = (1, 2, -4)$ in a zoomed-in portion of the grid of initial conditions used in \citet{Price-Whelan2016}. The white band in the top half of the figure represents a minor orbit family surrounding the commensurability (the structure in frequency-space shown here is equivalent to that shown in the bottom panel of Figure \ref{fig:01} for the Miyamoto-Nagai potential discussed earlier).}
\label{fig:05}
\end{figure}

Figure \ref{fig:05} shows a zoomed-in portion of the grid of orbits plotted in \citet{Price-Whelan2016} (compare to Figures 3, 4, and 5 in that paper). The zoomed-in map in Figure \ref{fig:05} is centered on the (1, 2, -4) commensurability, and the shading of each pixel represents the value of $\bm{n}\cdot\bm{\Omega}$. A large minor orbit family can be seen surrounding the commensurability (just as shown for the Miyamoto-Nagai potential in the lower panel of Figure \ref{fig:01}). The three points shown correspond to the initial conditions of the same types of orbits discussed in the previous sections, with A representing a commensurable orbit, B representing an orbit very close to the separatrix, and C representing an orbit that belongs to the major orbit family that envelopes the minor orbit family shown (in this case -- the long-axis tube orbit family). These three orbits are also three of the four orbits that were analyzed in \citet{Price-Whelan2016} (note that we have flipped the labels of orbits B and C from \citealt{Price-Whelan2016} to match the orbits discussed in the previous sections of this work).

As shown in \citet{Price-Whelan2016}, an ensemble initiated on orbit B forms a considerably more diffuse stream than those on orbits A or C. Given the proximity of the progenitor orbit to a separatrix, we consider the possibility that an ensemble initialized on this orbit may straddle the separatrix and therefore experience \textit{separatrix divergence}.

\begin{figure}
\includegraphics[width=0.85\columnwidth]{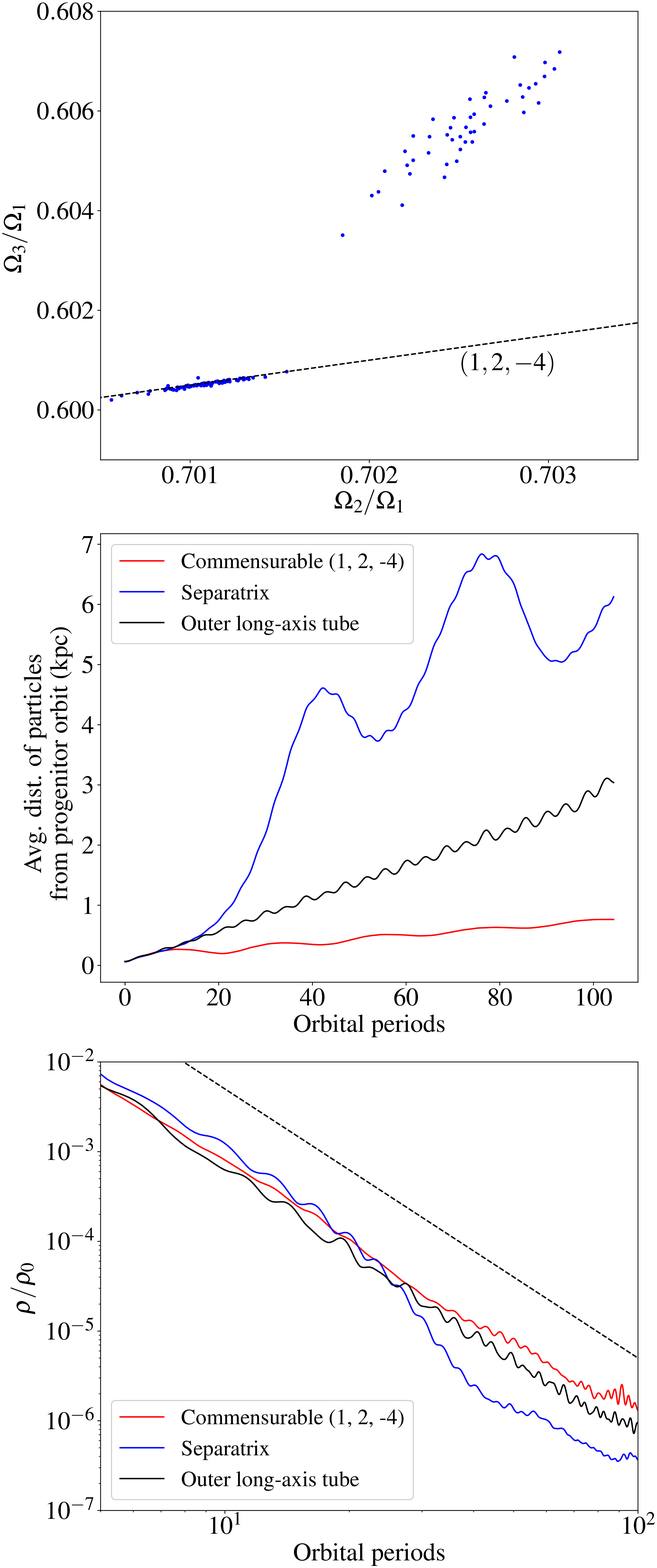} 
\caption{\textit{Upper panel}: frequency-frequency distribution of the ensemble whose progenitor is initialized on orbit B from Figure \ref{fig:05}. The (1, 2, -4) commensurability is shown with the dashed line. One subset of particles is trapped at this commensurability while the rest of the particles form a more diffuse yet distinct cluster far away from it. Compare this to the structure seen the lower-middle panel of Figure \ref{fig:02}. \textit{Middle and lower panels}: the mean distance of particles from the progenitor orbit and the mean density of the ensemble for each of the three orbits shown in Figure \ref{fig:05}. The expected $t^{-3}$ trend is plotted with a dashed line in the bottom panel.}
\label{fig:06}
\end{figure}

We begin by simulating three ensembles initialized on the orbits marked in Figure \ref{fig:05} following the initial conditions described in \citet{Price-Whelan2016}. The spread in the fundamental frequencies of ensemble B is shown in the top panel of Figure \ref{fig:06}, and demonstrates that the ensemble's particles fall neatly into two different groups, just like the frequency space plot shown for the separatrix ensemble in Figure \ref{fig:02} (the other two ensembles form one cohesive cluster in frequency space). 

Following the same steps as the previous section, we proceed by analyzing the mean distance of the ensemble particles from the progenitor orbit and the density evolution of the ensemble, shown in the middle and bottom panels of Figure \ref{fig:06}. The similarities between this figure and Figure \ref{fig:03} serve to further strengthen the notion that the reason for this ensemble's rapid diffusion in simulations is \textit{separatrix divergence}. In particular, the oscillatory behavior shown in the middle panel alongside the abrupt drop in density after $\sim30$ orbital periods shown in the bottom panel would be hard to explain under any other known mechanism for stream diffusion.

As in the last section, we compute the Lyapunov times for a representative subset of 64 particles. While seven of the 64 orbits returned Lyapunov times shorter than 500 orbital periods, the minimum Lyapunov time calculated for any of the orbits was $\sim300$ orbital periods -- still more than an order of magnitude longer than the timescale after which ensemble B appears to fan out. Moreover, the vast majority of orbits do not appear to be even mildly chaotic; the mildly chaotic behavior of $\sim10\%$ of the particles in this ensemble is insufficient to explain the rapid drop in density seen in the bottom panel of Figure \ref{fig:06}, nor does it justify the oscillatory behavior seen in the middle panel of the same figure. The lower panels of Figure \ref{fig:06} also serve to provide a rough estimate for the libration timescale of the commensurable orbit family ($T_l\sim100$ orbital periods).

The similarities between this case and the axisymmetric case discussed in \S\ref{sec:axisym} lead us to conclude that \textit{separatrix divergence} is very likely to be the main cause of this instance of stream-fanning.

\subsection{Stream-Fanning in the Law \& Majewski (2010) Model for the Milky Way}
\label{subsec:SP15}

We next turn to investigating the stream-fanning result in the \citet{Law2010} potential from \citet{Pearson2015}. The key finding from that paper was that a thin stream with the same characteristics as Pal 5 could not be reproduced with the correct position, velocity, and morphology in the \citet{Law2010} triaxial potential (whereas it could be reproduced to very high accuracy in a spherical potential). This led the authors to point out that stream morphology may be used as a method for ruling out certain shapes of the Galactic potential. The cause of the stream's rapid diffusion was left as an open question, though chaos was cited as a possible explanation.

Following \citet{Pearson2015}, we take the current position and velocity of Pal 5's progenitor orbit to be: $\bm{x} = (8.16,0.24,16.96)$ kpc and $\bm{v} = (56.43,101.23,3.84)$ km/s. We integrate this orbit backward in the \citet{Law2010} potential for 6 Gyr as a rough approximation of when Pal 5 began experiencing tidal stripping from the Milky Way.\footnote{\citet{Price-Whelan2019} showed that the previously measured distance to Pal 5 used by \citet{Pearson2015} was over-estimated. Furthermore, there is still much uncertainty around the time at which Pal 5 began experiencing tidal disruption, and a variety of integration times have been used in the recent literature to model Pal 5, ranging from under 3 Gyr \citep{Dehnen2004} to 10 Gyr \citep{Bovy2016}. However, since the objective here is to explain the stream-fanning effect in the \citet{Pearson2015} simulations, we adopt their initial conditions and integration time of 6 Gyr for a direct comparison to their results.}

\begin{figure}
\includegraphics[width=0.85\columnwidth]{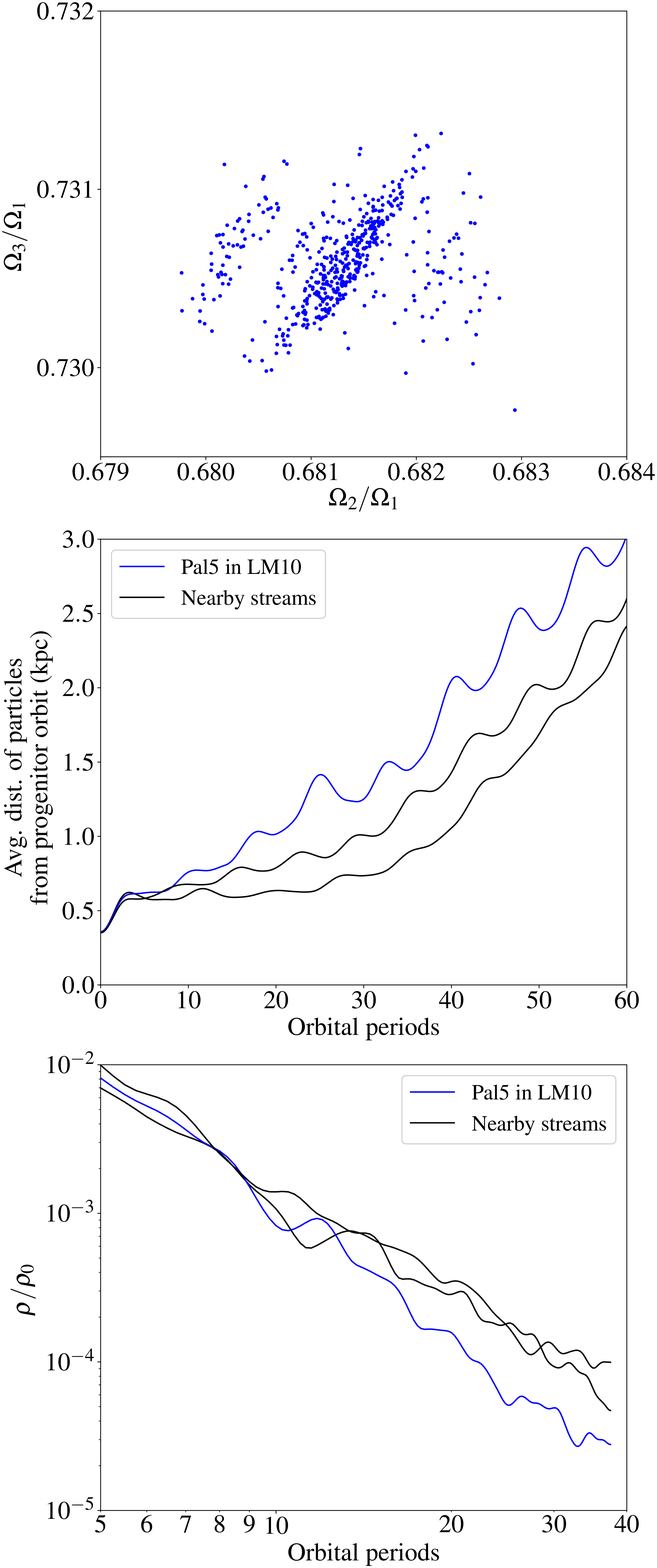} 
\caption{\textit{Upper panel}: frequency-frequency distribution of the ensemble whose progenitor is initialized on Pal 5's initial conditions based on \citet{Pearson2015}. Unlike the ensembles near the separatrices in the Miyamoto-Nagai potential in \S\ref{sec:axisym} or in the triaxial potential \S\ref{subsec:APW15}, here the particles do not fall neatly into two separate groupings. \textit{Middle and lower panels}: the mean distance of particles from the progenitor orbit and the mean density of the Pal 5 ensemble and two other ensembles initialized on nearby orbits.}
\label{fig:07}
\end{figure}

In order to model the evolution of Pal 5, \citet{Pearson2015} follow the \textit{Streakline} method described in \citet{Kupper2012}, whereas we continue to model the stream as an unbound ensemble of particles (using the same initial mass as \citet{Pearson2015}: $M=6\times10^4$ M$_\odot$). Like \citet{Pearson2015}, we find that an ensemble with Pal 5's initial conditions in the \citet{Law2010} potential no longer resembles a thin stellar stream after 6 Gyr. However, two ensembles initiated on nearby initial conditions (offset by $\pm1$ kpc along the $x$-axis) exhibit similar stream-fanning effects and do not remain considerably more stream-like than the ensemble with Pal 5's initial conditions. Unlike the previous two cases studied here, the frequency-frequency plot for the ensemble with Pal 5's initial conditions shown in the upper panel of Figure \ref{fig:07} does not show a clear tendency of the Pal 5 ensemble to split into two distinct groups.

The middle panel of Figure \ref{fig:07} shows that neither the Pal 5 ensemble nor the two ensembles we initialize near Pal 5's initial conditions display any hint of the oscillatory behavior induced by \textit{separatrix divergence} (such as the behavior shown in Figures \ref{fig:03} and \ref{fig:06}). The mean density evolution shown in the bottom panel of the same figure also reveals no clear tendency of the Pal 5 ensemble to diffuse faster or in a different manner than the nearby ensembles.

Unlike any of the other cases considered so far in this work, the \citet{Law2010} potential contains several components (a spherical bulge, an axisymmetric disk, and a triaxial halo that is misaligned with respect to the disk). The increased complexity of the potential gives rise to a more intricate web of commensurabilities that significantly complicate the analysis of the fundamental frequencies, and make it difficult to ascertain whether there is a single commensurability in close proximity to the progenitor orbit that dominates the evolution of the ensemble. Instead, the emerging picture from all three panels of Figure \ref{fig:07} is one in which several nearby commensurabilities each exert a varying influence on the orbit. While this could still be consistent with \textit{separatrix divergence}, it aligns more naturally with chaotic evolution of the stream, in which particles diffuse stochastically under the influence of multiple commensurabilities.

To test this hypothesis, we proceed by computing the Lyapunov times for a representative subset of 64 particles. In this case, we find that while the progenitor orbit has a Lyapunov time of $\sim20$ Gyr (as also reported in \citealt{Pearson2015}), many of the particles have even shorter Lyapunov times, and a few have Lyapunov times below $5$ Gyr. We therefore conclude that the cause for the stream-fanning of the Pal 5 ensemble in the \citet{Law2010} potential is more likely to be chaotic diffusion than \textit{separatrix divergence}.

\section{Limitations and Future Work}
\label{sec:discussion}

The main goal of this paper has been to describe a mechanism for stream-fanning that we label \textit{separatrix divergence}. We have focused on predicting when and where this mechanism will cause streams to fan out through a combination of test particle simulations, orbit family analysis, and the application of secular perturbation theory. We have made a series of simplifying assumptions in order to present our results, including working with smooth, static potentials and considering only the secular evolution of idealized streams made of unbound test particles. The gap between these conditions and the environment in which streams form around our Galaxy is wide, and includes many complicating factors that may either enhance or diminish the importance of \textit{separatrix divergence}, or change the morphological appearance of streams that have been affected by \textit{separatrix divergence}.

In the next few paragraphs, we discuss some of these shortcomings in greater depth, alongside some promising avenues for how this analysis might be extended to overcome these drawbacks and provide meaningful constraints on the shape of our Galaxy's dark halo.

\subsection{Accurate Modelling of Stellar Streams}
\label{subsec:stream_modelling}

The findings above rest on the assumption that stellar streams are well-modelled by an unbound ensemble of massless particles whose initial conditions are spread in an orderly fashion around the progenitor orbit. In reality, a stream forms gradually as the progenitor orbits and its stars are tidally stripped by the host Galaxy. Since the progenitor is not, in fact, massless, it will continue to affect the orbits of nearby stars even after they have been tidally stripped. Furthermore, the progenitor's gravitational influence may also distort the local orbital structure of the potential. Such distortions could make it less likely for stars to be deposited in a different orbit family than that of the progenitor's orbit.

Even assuming the previous point is relatively unimportant, the gradual nature of the stripping process leads to a qualitatively different picture of \textit{separatrix divergence} than the one described throughout this work. Instead of the entire ensemble fanning out upon reaching the unstable (hyperbolic) fixed point for the first time, only the stars that have been tidally stripped at that point should exhibit this diffusion. Stars that are stripped subsequent to the first passage of the hyperbolic fixed point will evolve as a thin stream, and will only fan out after the second passage of the fixed point. As a result, one may expect the stream to appear at any given time as a thin stream that is truncated at a certain point, beyond which all of the particles are fanned-out.

\subsection{Observability of \textit{Separatrix Divergence}}
\label{subsec:observability}

A key drawback of \textit{separatrix divergence} and using it to understand our Galaxy's potential rests in that it predicts more diffuse stream morphologies that would therefore be \textit{harder} to detect. However, outside of the possibility of detecting direct evidence of \textit{separatrix divergence} despite the lower surface brightness of the fanned-out morphologies, we posit two characteristics of \textit{separatrix divergence} that may be exploited to identify a stream in the vicinity of a separatrix:
\begin{enumerate}
	\item Throughout this work we have described the outcome of \textit{separatrix divergence} as stream-fanning, in which all of the stream's stars end up forming a more diffuse structure. In practice, several more intriguing morphologies are also theoretically possible. In particular, the formation of two distinct groupings in frequency space (see Figures \ref{fig:02} and \ref{fig:06}) raises the possibility that this may lead to a detectable feature like a bifurcation.
	\item The qualitative picture described in \S\ref{subsec:stream_modelling} offers another promising solution: one may hope to detect \textit{separatrix divergence} by locating a tidally disrupted globular cluster with truncated streams. If a diffuse collection of stars at the edge of the stream can be associated with the rest of the stream stars through either their orbital properties of their chemistry, it may be possible to associate this truncation with \textit{separatrix divergence}.
\end{enumerate}

\subsection{Potential Mapping and Constraining the Milky Way's Shape}
\label{subsec:potential_mapping}

Throughout this work we have focused on describing the theoretical basis for \textit{separatrix divergence} and its morphological effects on stellar streams. Further progress necessitates an ability to map the orbital structure of a potential and predict locations where \textit{separatrix divergence} may occur.

A preliminary step toward potential mapping that incorporates the effects of \textit{separatrix divergence} is shown in Figure \ref{fig:05} for the Miyamoto-Nagai axisymmetric potential. Yet in practice one is likely to need a map that covers more than one fixed set of values for $E$ and $L_z$, not to mention the fact that relying on surfaces of section automatically rules out the ability to extend this specific map to a 3-D non-axisymmetric potential.

The ability to produce potential maps that specify the location and likelihood of \textit{separatrix divergence} would enable two new approaches for constraining a potential. First, the detection of an individual stream whose morphology indicates that it has been affected by \textit{separatrix divergence} could provide a very strong constraint by pinpointing the precise location of a boundary between two orbit families. Aside from indicating that the potential is able to support multiple orbit families, the exact location of a commensurability should provide significant constraining power. Of course, one would first have to rule out other mechanisms that could have caused the observed stream morphology.

Alternatively, one could approach this question from a statistical perspective. Even if \textit{separatrix divergence} renders streams completely undetectable, there may still be information contained in the number and distribution of observable thin streams, in the sense that one can rule out configurations that would cause the streams to fan out. One might extend this to ask questions like: what is the probability that the potential has a certain shape given the number of thin streams detected? Are there certain areas in phase-space that have a suspicious deficiency in thin streams, possibly indicating that all of the streams in that neighborhood have become too diffuse to detect due to \textit{separatrix divergence}?

\section{Summary and Conclusion}
\label{sec:conclusion}

In this work, we have described a mechanism for the fanning of stellar ensembles in galactic potentials that support multiple orbit families. The separatrices between those orbit families represent discontinuities in the orbital structure of the potential, and the characteristics of orbits on either side of a separatrix can differ greatly. The tidal disruption of a globular cluster whose orbit is close to a separatrix may deposit stars on both sides of the separatrix, leading us to describe this ensemble as `straddling' the separatrix.

By comparing test particle simulations with analytically derived quantities from perturbation theory, we have shown that a stream born in this configuration is susceptible to rapid diffusion (which we name \textit{separatrix divergence}) within a time that can be much shorter than the stream's typical phase-mixing timescale. We describe two diagnostics that can be used to relate between the stream's initial conditions and its likelihood of experiencing \textit{separatrix divergence}:
\begin{enumerate}
	\item The timescale after which an ensemble will fan out due to \textit{separatrix divergence} is directly related to the libration frequency of the commensurable orbit family, which can be calculated analytically using Hamiltonian perturbation theory (see equation \ref{eq:libration_freq}). If this timescale is shorter than the typical phase mixing timescale of the stream (obtained from equation \ref{eq:hessian}), \textit{separatrix divergence} will have a significant effect on the stream's morphological evolution.
	\item The strength of the effect can be estimated by comparing the extent of the discontinuity in action-space at the separatrix (see equation \ref{eq:libration_action}) with the initial spread of the stream in action space (equation \ref{eq:action-spread}). If the former is greater than the latter, \textit{separatrix divergence} will cause a stream to become considerably more diffuse than similar streams that do not encounter a separatrix.
\end{enumerate}

We utilized test particle simulations to demonstrate these effects for an axisymmetric potential. Next, we analyzed two reports from previous literature of stream-fanning in simulations of triaxial potentials, concluding that at least one of them is likely to have been the result of \textit{separatrix divergence}. 

Our results suggest that the morphology of stellar ensembles may be able to constrain the shape of a galaxy's gravitational potential. We are optimstic that the findings of this work can be used to map out a potential in terms of regions that will or will not support the formation of thin streams. Used alongside observations of stellar streams, this technique offers a new tool for validating or ruling out models for the gravitational potential of our Galaxy.

\section*{Acknowledgements}

It is a pleasure to thank David Hendel, Greg Bryan, David Helfand, the \textit{Stream Team} and the \textit{Milky Way Stars} group at Columbia University, and the Dynamics Group at the Center for Computational Astrophysics, for invaluable comments, discussions, and advice. We are grateful to the anonymous referee whose detailed and thoughtful comments provided many insights that greatly improved the manuscript.

TDY is supported through the NSF Graduate Research Fellowship (DGE-1644869). TDY and KVJ were partially supported by the National Science Foundation under Grant No. AST-1715582.

We thank the Center for Computational Astrophysics at the Flatiron Institute for support and space to conduct this work.

This research made extensive use of AGAMA \citep{Vasiliev2019}, Astropy \citep{Robitaille2013,Price-Whelan2018b}, Gala \citep{Price-whelan2017}, Galpy \citep{Bovy2014a}, Matplotlib \citep{Hunter2007}, Numpy \citep{VanDerWalt2011}, Scikit-Learn \citep{scikit-learn}, Scipy \citep{Virtanen2019}, Superfreq \citep{Price-Whelan2015}, and TorusMapper \citep{Binney2016a}.

\section*{Data Availability}

The data underlying this article will be shared on request to the corresponding author.


\bibliographystyle{mnras}
\bibliography{mybib}




\bsp	
\label{lastpage}
\end{document}